\documentclass{article}
   \usepackage{graphicx}
   \usepackage{amssymb}
   \usepackage{epstopdf}
   \usepackage{apjgalley}
   \usepackage{multicol}
\font\sevenrm=cmr7  
\def\teq#1{$\, #1\,$}                         % text equation
\def\tsyn{t_{\hbox{\sevenrm syn}}}
\def\tacc{t_{\hbox{\sevenrm acc}}}
\def\gammax{\gamma_{\hbox{\sevenrm max}}}
\def\Ebreak{E_{\hbox{\sevenrm br}}}

\begin{document}

\newcommand{\figureoutpdf}[5]{\centerline{}
   \centerline{\hspace{#3in} \includegraphics[width=#2truein]{#1}}
   \vspace{#4truein} \figcaption{#5} \centerline{} }
\newcommand{\twofigureoutpdf}[3]{\centerline{}
   \centerline{\includegraphics[width=3.4truein]{#1}
        \hspace{0.5truein} \includegraphics[width=3.4truein]{#2}}
        \vspace{-0.2truein}
    \figcaption{#3} }    % plot side-by-side PDF version for TeXShop
\newcommand{\twofigureoutoverpdf}[3]{\centerline{}
   \centerline{\includegraphics[width=3.85truein]{#1}}
        \vspace{0.0truein} 
   \centerline{\includegraphics[width=3.85truein]{#2}}
        \vspace{0.0truein}
    \figcaption{#3} }    % plot on-top-of-other PDF version for TeXShop
\newcommand{\twofigureoutcustompdf}[3]{\centerline{}
   \centerline{\includegraphics[width=2.63truein, angle=270]{#1}}
        \vspace{0.0truein} 
   \centerline{\hspace{0.02truein} \includegraphics[width=3.51truein]{#2}}
        \vspace{0.0truein}
    \figcaption{#3} }    % plot on-top-of-other PDF version for TeXShop

\title{A Suzaku X-Ray Study of the Particle Acceleration Processes in the Relativistic Jet of Blazar Mrk 421}

   \author{Alfred B. Garson III} 
   \affil{Washington University in St. Louis \\
       and The McDonnell Center for the Space Sciences, St. Louis, MO 63130\\
       \it agarson3@hbar.wustl.edu\rm}

   \bigskip
    
   \author{Matthew G. Baring}
   \affil{Department of Physics and Astronomy MS-108, \\
      Rice University, P.O. Box 1892, Houston, TX 77251, U.S.A.\\
      \it baring@rice.edu\rm}

 \and

   \author{Henric Krawczynski}
   \affil{Washington University in St. Louis \\
      and The McDonnell Center for the Space Sciences, St. Louis, MO 63130\\
      \it krawcz@wuphys.wustl.edu\rm}

\slugcomment{Accepted for publication in \it The Astrophysical Journal\rm}

\begin{abstract}
We report on the findings of a 364 ksec observation of the BL LAC object
Mrk 421 with the X-ray observatory Suzaku. The analysis in this paper
uses fluxes and hardness ratios in the broad energy range from 0.5 keV
to 30 keV. During the course of the observation, the 0.5 keV - 30 keV
flux decreased by a factor of $\sim$2 and was accompanied by several
large flares occurring on timescales of a few hours. We find that
fitting a broken power model to spectra from isolated epochs during the
observation describes the data well. Different flares exhibit different
spectral and hardness ratio evolutions. The cumulative observational
evidence indicates that the particle acceleration mechanism in the Mrk
421 jet produces electron energy distributions with a modest range of
spectral indices and maximum energies.  We argue that the
short-timescale X-ray spectral variability in the flares can be
attributed {\it mostly} to intrinsic changes in the acceleration
process, dominating other influences such as fluctuations in the Doppler
beaming factor, or radiative cooling in or outside the acceleration zone.
\end{abstract}

\keywords{BL Lac, blazars, jets, X-ray: general --- blazars, jets, X-ray: individual (Mrk 421)}

\section{Introduction}
 \label{sec:intro}
TeV blazars exhibit $\nu$F$_{\nu}$ spectral energy distributions (SEDs)
with two broad  peaks: one in soft to medium X-rays and one at GeV
energies.  The amplitude and the position of the peak changes with time,
sometimes in a correlated way (\cite{BL05}). The emission is highly
polarized in radio through optical wavelengths (\cite{PE05,HO00,LS00}).
The radio through X-ray spectrum is thought to be the result of
synchrotron emission from the highest energy electrons and positrons
accelerated to Lorentz values $\gamma_e\gg10^{2}$ by shock fronts in the
jet.  Inverse Compton scattering from the same population of electrons
and their synchrotron photons may be responsible for the peak at higher
energies. This synchrotron self-Compton (SSC) model is contrasted by
external Compton (EC) models. In the latter the low-energy photons
originate outside the emission volume of the gamma-rays. Possible
sources of target photons include: accretion disk photons radiated
directly into the jet, accretion disk photons scattered by emission-line
clouds or dust in the jet, synchrotron radiation re-scattered back into
the jet by broad-line emission clouds, jet emission from an outer slow
jet sheet, or emission from faster or slower portions of the jet
(\cite{GM89,Ma95,Mk97,Gh05}, \cite{Ge04}).

There are also hadronic models for the TeV emission. An example for a
hadronic $\gamma$-ray production mechanism is pion photoproduction from
either low energy synchrotron photons or photons external to the jet
(\cite{MA93,MU03}). Synchrotron emission from protons in compact regions
of the jet is another explanation (\cite{AR00}).

Blazars are known for their variability at X-ray and $\gamma$-ray
energies. The X-ray and $\gamma$-ray  fluxes can vary rapidly and are
often correlated, with a notable exception of orphan TeV flares
(\cite{KR04}). X-ray flaring epochs lasting many months have been observed
as well as sub-hour flares (\cite{CU04}). The source of flaring activity
has been attributed to internal shocks within the jet (\cite{RE78,SP01}),
ejection of relativistic plasma into the jet (\cite{BO97,Mk97}), as well
as reconnection events in a magnetically dominated jet (\cite{LY03,DD09}).

Constraining blazar jet models generally requires simultaneous
observations across the radio to gamma-ray spectrum. However, these
probes can be augmented by focusing on the nuances in a particular
waveband.  Such is the approach here, specifically with X-rays and their
variable signals, since the high count rates and good spectral
resolution in this band provide powerful additional probes of the jet
environment.  Note that using the X-ray spectra of TeV blazars can
provide constraints on the modeling of the overall SED. The position of
the synchrotron peak has become a marker for classes of BL Lac objects
(\cite{PG95}). LBL (Low energy peaked BL Lac) and HBL (high energy peaked
BL Lac) designate whether the synchrotron peak in is the IR-optical or
UV-X-ray bands, respectively. Monitoring spectral parameters as flaring
events evolve gives insight into the mechanics of the emission
(\cite{KI98}). From their analysis, for simple models involving a single
electron population, different hardness ratio (HR) trends will be
observed for varying fluxes depending on the timescales of the processes
involved. In the HR-flux plane, there will be clockwise movement as time
progresses if the high energy component varies faster than the low
energy component, where electron cooling times exceed the acceleration
time; this case is more probably sampled below the X-ray band.
Counter-clockwise motion in the HR-flux plane is predicted if the
observation is made near the synchrotron cutoff frequency, specifically
when the cooling and acceleration timescales are roughly equal.

In the case of TeV bright blazars like Mrk 421, individual sources have
shown both hard and soft lags (\cite{KA00,TA00,SA08}). Such lags are
apparent when a light curve is examined in two energy bands (canonically
separated at 2 keV). Trends in the count rate do not always occur
simultaneously in both energy windows. Features can also be observed
first at high energies (soft lag) or first at low energies (hard lag).

The peak energy and curvature of the X-ray spectrum have been shown to
be anti-correlated for different acceleration scenarios such as
stochastic or energy dependent acceleration (\cite{KA62,MA04,SP08}). 
X-ray measurements also can provide limits on physical properties of the
emitting region such as its size (e.g. R$\approx$10$^{14}$cm; see
Tramacere et al. 2009). Thus, although multi-wavelength observations are
crucial for investigating acceleration and emission processes, careful
study of X-ray observations from TeV blazers can give insight into the
mechanisms responsible for the populations of charged particles and
photons in the jet.

Mrk 421 is a TeV blazar and, at a redshift of z=0.031, it is one of the
closest and best-studied BL Lac objects. It was the first extra-galactic
TeV source (\cite{PU92}) and has been the target of many multiwavelength
campaigns (\cite{TA96,KR01,RE06,GU08,FA08}). The synchrotron peak in Mrk
421's spectrum ranges from a fraction of a keV to several keV and
spectral variability as a function of flux level has been observed
(\cite{FA00}). In general, the spectrum becomes harder for higher fluxes,
both in the X-ray band (e.g. Fossati et al. 2008; Tramacere et al. 2009)
and in the gamma-ray regime (\cite{KR02,AR02}). It also now has a
well-measured GeV-band spectrum from {\it Fermi}'s Large Area Telescope
(see Abdo et al. 2009) that provides useful constraints on the high
energy electron population using an inverse Compton signal
interpretation. The relationship between the {\it Fermi} and
\textit{Suzaku} spectra will be discussed in Section~\ref{sec:discuss}
below.

Takahashi et. al (1996) observed a soft lag ($<$1.5 keV) in X-rays. When
attributed to synchrotron electron lifetimes, the magnetic field
strength and electron Lorentz factor were found to be B $\sim$0.2 G and
$\gamma_e\sim$10$^{6}$, respectively. \textit{Swift} observations
indicate that each flare has it's own competition between time scales
involved with electron acceleration and cooling. The energy spectrum of
the electrons associated with the UV-X-ray emission can be described
with a curved population (\cite{TR07,TR09}).  Previous \textit{Suzaku}
observations suggest that the emmission contains a steady component and
a variable component. The latter may be attributed to localized Fermi I
type acceleration in individual shocks, while the former may be due to
superposition of shocks at larger distances from the jet or other
processes (\cite{US09}).

In this paper, we give the findings from a 4-day observation of Mrk 421
with the X-ray satellite telescope \textit{Suzaku} in May 2008. This
pre-dates the launch of Fermi. Simultaneous XMM-Newton and VERITAS
gamma-ray observations in a separate campaign were described by Acciari
et al. (2009).  Another paper combines a large number of multiwavelength
observations of Mrk 421, including the XMM Newton, \textit{Suzaku} and
VERITAS data (Acciari et al. 2010).

We investigate the evolution of spectral parameters over the duration of
the observation.  The study presented in this paper benefits from the
long exposure of 364 ksec and the excellent sensitivity of
\textit{Suzaku} over the 0.5 keV to 30 keV energy range. Compared to the
2006 \textit{Suzaku} observation campaign presented by Ushio et al.
(2009), the observations presented here reveal the source in a lower
flux state. In Section~\ref{sec:observe} we describe the \textit{Suzaku}
intruments, give the essentials of this observation, and outline the
analysis protocol.  The results are detailed in
Section~\ref{sec:analysis}.  These are followed in
Section~\ref{sec:discuss} by the discussion of  the interpretation and
implications of the findings, highlighting how the X-ray spectrum and
variability impacts our understanding of the Mrk 421 jet environment and
the particle acceleration properties therein.

\section{Observations and Data Reduction}
 \label{sec:observe}
\textit{Suzaku} (\cite{MI07}) is an X-ray observatory with two primary
instruments. The X-ray Imaging Spectrometer (XIS) is an imaging X-ray
CCD instrument with 3 operating detectors: two are sensitive from 0.5
keV to 10.0 keV (XIS0 and XIS3), while the backside-illuminated XIS1
extends the low energy range to 0.2 keV (\cite{KO07}). Complementary to
and co-aligned with the XIS is the Hard X-ray Detector (HXD) which is a
well-type instrument composed of GSO scintillator and silicon PIN
diodes. The PIN detectors observe in the 12 keV to 60 keV energy band,
while the GSO can detect up to gamma-ray energies (\cite{TA07,KK07}). This
observation (ID 703043010) was triggered from a detection by the
ground-based atmospheric \v{C}erenkov telescope, VERITAS. Mrk 421 was
observed May 5 2008 02:52 (MJD 54591) through May 9 08:24 (MJD 54595).
\textit{Suzaku} has two observation modes which place a source either in
the center of the HXD or XIS fields of view (FOV). HXD pointing was
selected for this observation. The XIS instruments were operated in 1/8
window mode.

\subsection{Data Reduction}
 \label{sec:reduction}
The XIS and HXD event files were used for this study. Standard reduction
and processing were performed using HEASOFT v6.8 and \textit{Suzaku}
ftools v15. The files were cleaned with the selection criteria: cutoff
rigidity larger than 6 GV/c, Earth rim elevation angle greater than
5$^{\circ}$ and 20$^{\circ}$ during the night and day, respectively.

XIS events were extracted from a source region with an inner radius of
35 pixels and an outer radius of 408 pixels. The extent of the inner
radius is such that pile-up effects were minimized for the selected
events. The background was selected from an annulus outside of the
source region, with inner and outer radii of 432 pixels and 464 pixels,
respectively. The response matrix and effective area were calculated for
each XIS sensor using the \textit{Suzaku} ftools tasks,
\textit{xisrmfgen} and \textit{xissimarfgen} (\cite{IS07}). XIS1 data were
not included in this analysis; including the XIS1 spectra did not
improve the quality of the fits. As the XIS0 and XIS3 have similar
reponses, their data were summed.

PIN data were extracted from the HXD uncleaned event files after
standard screening. The tuned background model supplied by the
\textit{Suzaku} team was used for Non X-ray Background (NXB) events. 
The source spectra were corrected for deadtime using \textit{hxddtcor}. The
PIN light curves were deadtime-corrected bin-by-bin (after incorporating
4 ksec - 6 ksec bins) using pseudo events generated in orbit. The
background and spectra light curves were corrected for their 10x
oversampling rate. We estimate the cosmic X-ray background (CXB)
contribution to the PIN background using the model given in Gruber et
al. (1999), which is folded with the PIN response to estimate the CXB rate.

\section{Temporal and Spectral Results}
 \label{sec:analysis}
\subsection{Light Curves}
We plot the time history of the observation in Figure \ref{LC}. Light
curves are given in 2 energy bands for each instrument: 0.5 keV - 2.0
keV and 2.0 keV - 10.0 keV for the XIS; 10.0 keV - 20 keV and 20 keV -
30 keV for the PIN. The three lowest energy bands include similar count
rate evolution throughout the observation, while the highest energy band
does not have significant changes. Overall, the rates decrease by up to
a factor of 2 over the course of the observation. Rates decrease from
$\sim$50 cts/sec to $\sim$20 cts/sec in the 0.5 keV - 2 keV range and
0.16 cts/sec to 0.13 cts/sec in the 20 keV - 30 keV band. The general
decline in rates is marked by several shorter duration flares occurring
at 80ks, 120ks, 140ks, 260ks, 310ks and 340ks after the start of the
observation.

\figureoutpdf{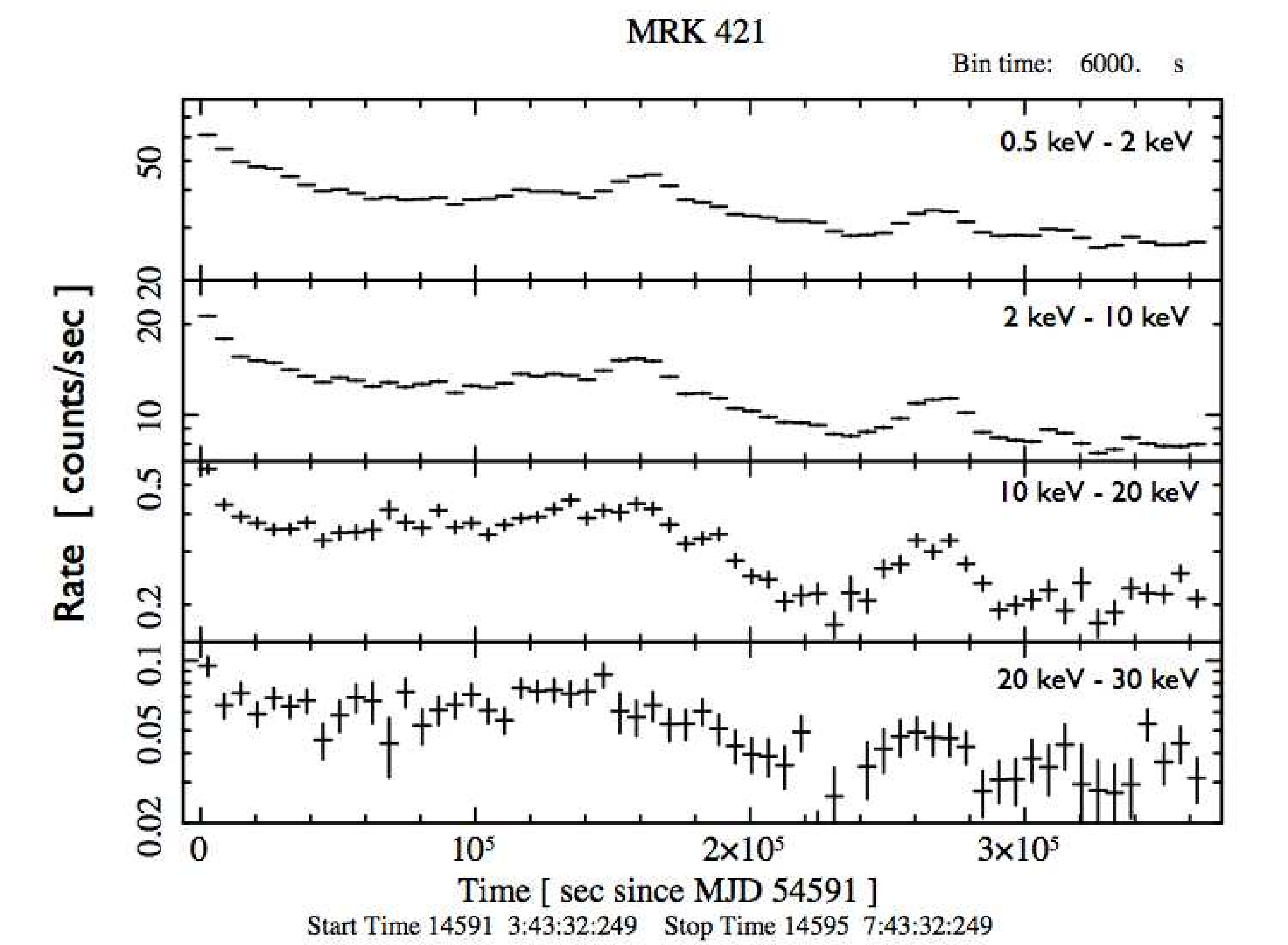}{3.3}{0.0}{-0.1}{
%\begin{figure}
%\includegraphics[scale=.6]{figs/fig1.eps}
%\caption{
These light curves illustrate the rates measured by the 
XIS and PIN detectors. Curves are given for (\textit{top to bottom}) 
0.5 keV - 2 keV, 2 keV - 10 keV, 10 keV - 20 keV, and 20 keV - 30 keV. 
Over the course of the observation, the rates decrease by a factor 
of $\sim$2. The light curve is marked with 2 strong flares and several 
small flares. \label{LC}}
%\end{figure}

The light curves begin with a rapid decline in rates for the initial
20ks of the observation. There is a general leveling off with some small
variations in rates over the next 100ks. A strong $\sim$25ks flare is
then seen which brings rates close to their original level. This flare
is shorter in duration for higher energy bands. The rates then smoothly
decline for $\sim$30ks. During this time, a flare is observed in the 12
keV - 20 keV band, but not in any other bands. There is then a second
large flare which is seen in the three lower energy ranges. The last
50ks of the observation has two small flares spanning the period.

We investigate the time evolution of the hardness ratio (HR) of the XIS
events.  It is convenient to divide the observational window into two
bands, $0.5 - 2$keV(\textit{a}) and $2 - 10$keV (\textit{b}), and define
the HR as either the ratio of counts (\textit{b/a}), or the ratio of the
difference and sum of counts (\textit{(b-a)/(b+a)}), a standard
protocol. Here we use the former definition of HR. This differs slightly
from the approach of Tramacere et al. (2009), who use the spectral index
at 1 keV to prescribe a hardness ratio.

Figure \ref{LCHR} shows the 0.5 keV - 2.0 keV (\textit{top}) and 2 keV -
10 keV (\textit{middle}) rates, and the corresponding hardness ratio
(\textit{bottom}). At the start of the observation, the rates decline
quickly as does the HR. At a time of $\sim$20ks, the HR begins to
increase while the rates in both bands continue to decrease then
stabilize for a duration of $\sim8000$ sec. For the remainder of the
observation, the HR follows the flux. It becomes harder for larger
fluxes so that the HR vs time plot largely reproduces the features in
the light curves. Using the XIS response and xspec, we simulate spectra
for a simple power-law model using photon indices between 2.2 and 2.5.
We calculate the hardness ratio measured  by XIS for these simulated
observations and indicate the corresponding position in the lower panel
of Figure \ref{LCHR} (\textit{horizontal dotted lines}) for comparison
with the measured HRs.

\figureoutpdf{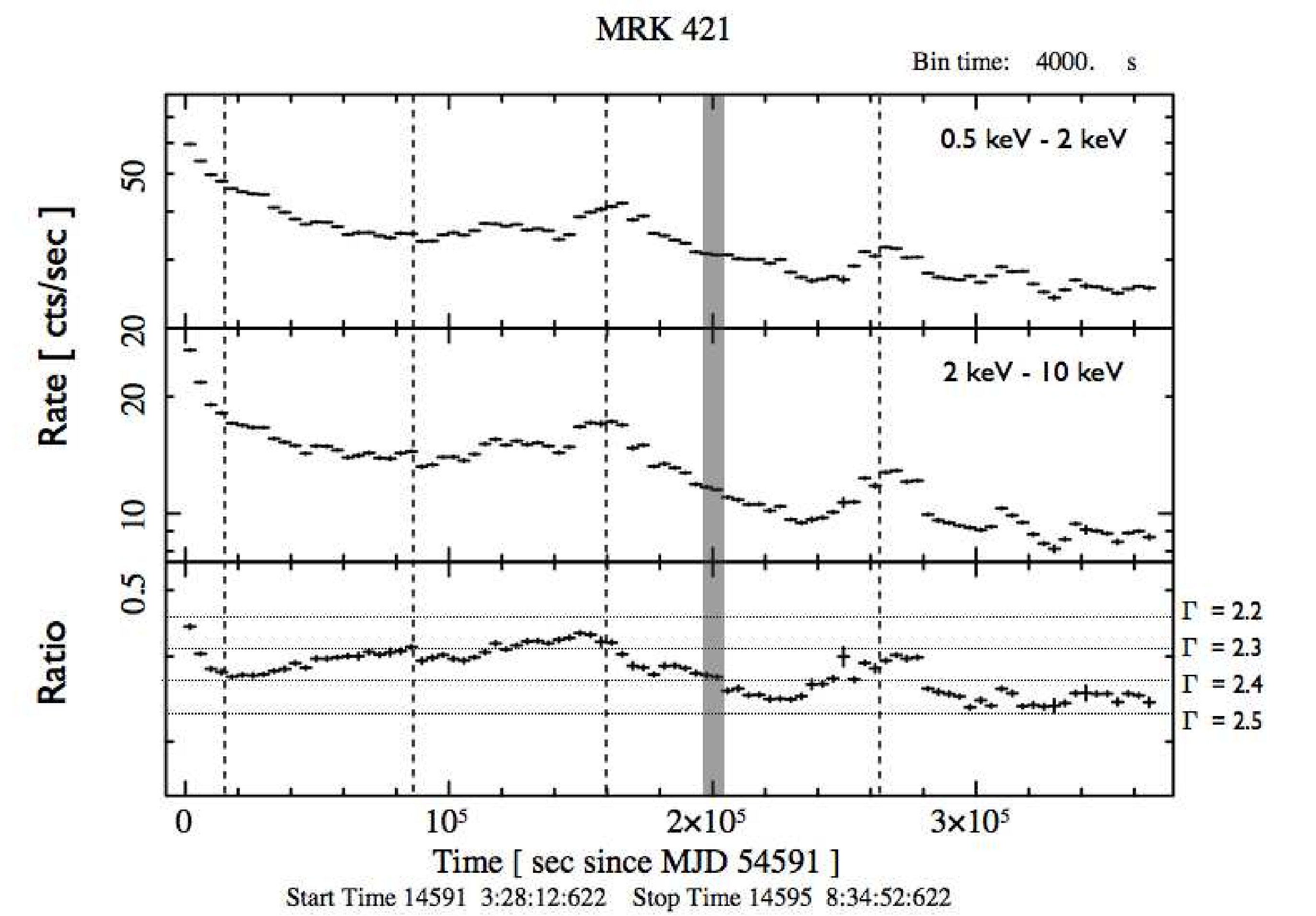}{3.3}{0.0}{-0.1}{
%\begin{figure}
%\includegraphics[scale=0.6]{figs/fig2.eps}
%\caption{
This light curve samples the XIS count rate with slightly finer 
binning (4000s) than in Figure \ref{LC} (6000s). The high and low 
energy bands are given in the top and middle panels, respectively. 
The bottom panel shows the corresponding hardness ratio (HR=rate
$_{High}$/rate$_{Low}$). Using the XIS response files, pure power-law 
spectra observations were simulated. The horizontal dotted lines in the 
lower panel give expected hardness ratios using simulated power-law 
spectra with a range of power-law photon indices. The shaded region 
near 2$\times$10$^{5}$ sec indicates the coincident VERITAS observation 
window from Acciari et al. (2009).
 \label{LCHR}}
%\end{figure}

The vertical lines in Figure \ref{LCHR} are to aid the eye in comparing
trends between the rates in the two energy bands and the HR. The first
vertical line indicates when the HR changes from decreasing to
increasing trend while the count rates continue to decrease. The second
vertical line shows a time when the rates and the HR level off and drop
during a small flare. The third vertical line marks the peak of a large
flare (\textit{Flare 1}) in both energy bands. However it is obvious
that the HR peaked $\sim$20 ksec prior to the peak in rates. The last
vertical line is again placed at a peak  (\textit{Flare 2}) in the count
rates. For this flare, the HR peak is located closer in time to the peak
rate, however the subsequent decrease in the HR is delayed compared to
the rate decrease. The shaded gray region in Figure~2 denotes the
observation window for the VERITAS campaign described in Acciari et al.
2009; see also Acciari et al. 2010).

\newpage

\subsection{Spectra}
During fitting, events with deposited energy between 1.5 keV and 2.5 keV
were excluded from the XIS data set due to uncertainties in the
instrument response (\cite{US09}). Events with energy between 10 keV and
25 keV were included from PIN data. The resulting spectrum was fit with 
a galactic absorption $\times$ broken powerlaw . The galactic absorption
parameter, $n_H$, was kept constant at the value of 1.61$\times 10^{20}$
cm$^{-2}$ acquired from the CIAO tool
Colden\footnote{http://cxc.harvard.edu/toolkit/colden.jsp}. Once the
normalization parameter was fit for the model, it was frozen while the
low energy photon index  ($\Gamma_{1}$), break energy ($\Ebreak$), and
high energy photon index ($\Gamma_{2}$)  were fit independently.
Finally, all parameters were simulataneously fit. Note that while
similarly good spectral fits were produced using power-law with
exponential cutoff models for a few of the results presented here,
broken power-law fits were superior to other spectral functions for the
majority of time intervals. For broken power-law fits, the average
reduced $\chi^2$ is 1.1 for 41 degrees of freedom (dof) with a standard
deviation of 0.34, while power-law with exponential cutoff fits produced
average reduced $\chi^2$ of 4.05 for 42 dof with a standard deviation 
of 2.1.

\twofigureoutcustompdf{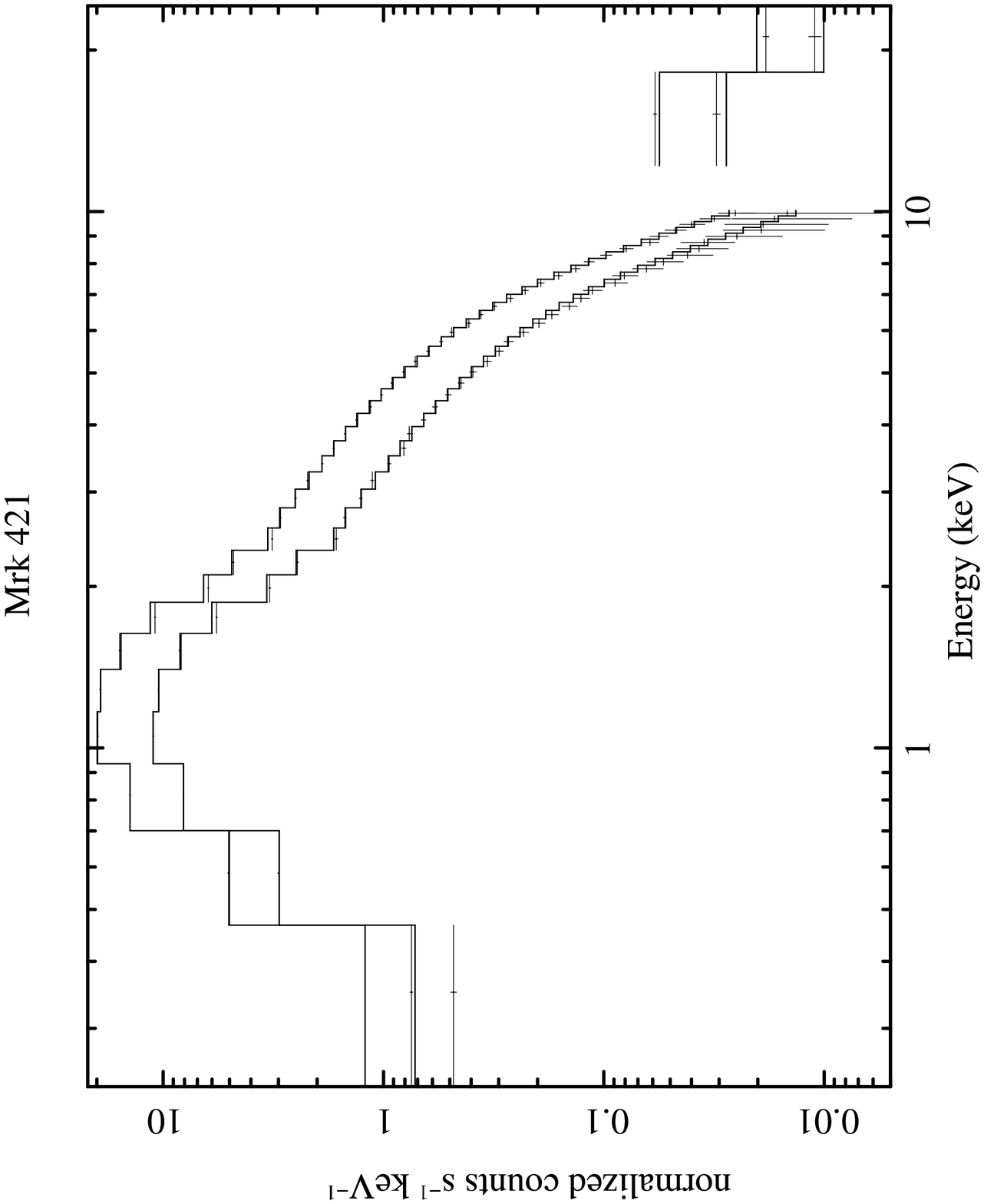}{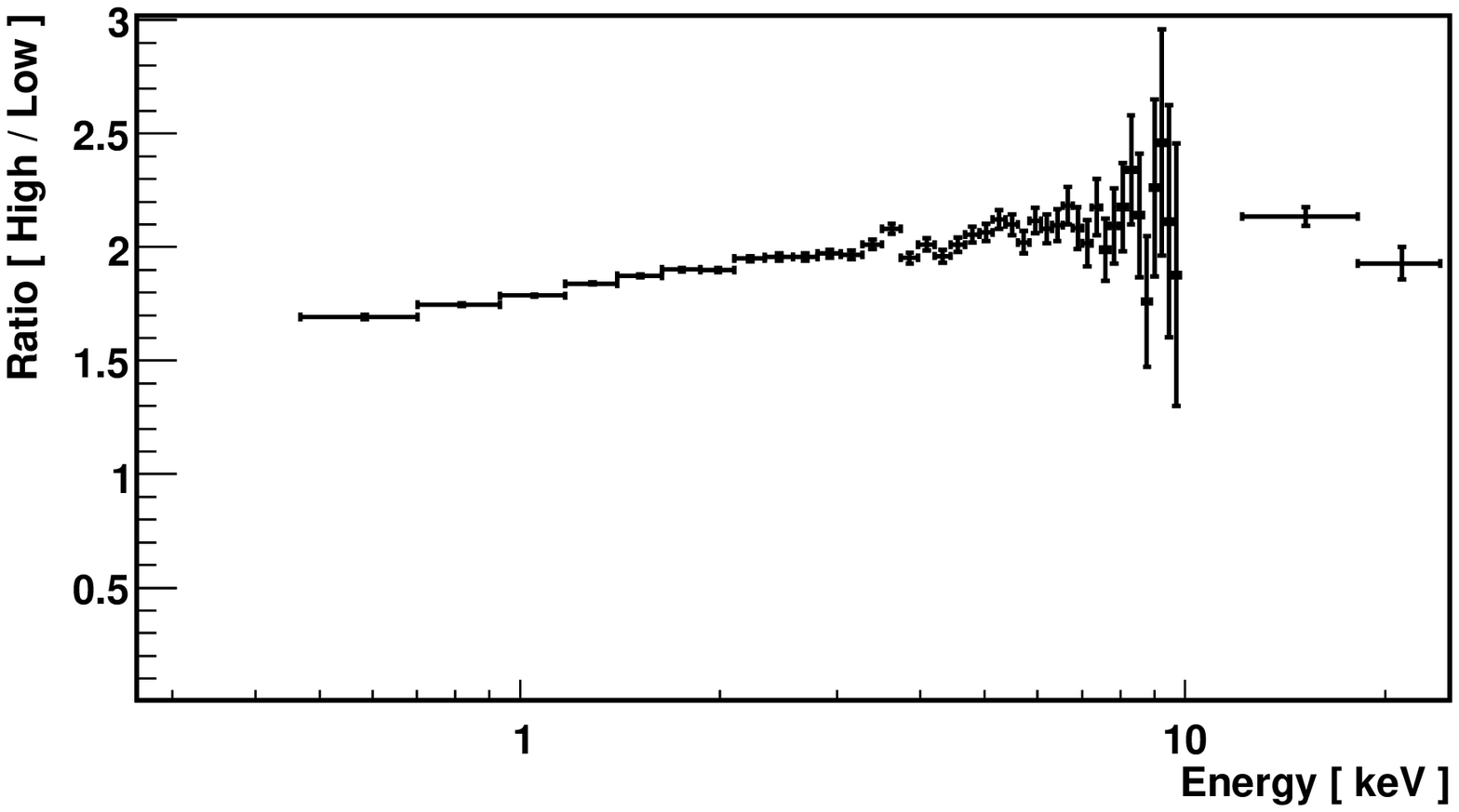}{
%\begin{figure}
%\includegraphics[scale=0.51, angle=270]{figs/fig3a.eps}\\
%\includegraphics[scale=0.67,]{figs/fig3b.eps}
%\caption{
The XIS and PIN observed spectra are given for the first 
$\sim$20 ks (\textit{High}) and the final $\sim$20 ks (\textit{Low}) 
of the observation (\textit{top panel}). The bottom panel plots the ratio, 
Rate$_{High}$/Rate$_{Low}$, as a function of energy. A factor of $\sim$2 
count rate decrease can be seen at most energies between 3 keV and 
20 keV while smaller ratios occur at lower energies. \label{FILA}}
%\end{figure}

Comparing the spectra at the endpoints of the observation can give
insight into the overall evolution of the emission as a function of
energy. The upper panel of Figure \ref{FILA} plots the XIS and PIN
spectra for the first as well as final $\sim$20ks of the observation.
The best fit broken power-law models are also shown. The best fit
parameters for the start of the observation are $\Gamma_{1}$=2.31 
$\pm$0.007,  $\Gamma_{2}$=2.6  $\pm$0.008,  E$_{B}$=2.8 $\pm$0.075 keV,
and normalization of 0.37  $\pm$0.0006 cts s$^{-1}$ keV$^{-1}$ producing
a reduced $\chi^{2}$ of 1.774.  The best fit parameters for the end of
the observation are $\Gamma_{1}$=2.39  $\pm$0.03,  $\Gamma_{2}$=2.62 
$\pm$0.01,  E$_{B}$=2.37 $\pm$0.3 keV, and normalization of 0.2 
$\pm$0.0008 cts s$^{-1}$ keV$^{-1}$ producing a reduced $\chi^{2}$ of
1.19.  It is apparent that the rates do decline over the course of the
observation. The lower panel shows the hardness ratio, (Rate$_{High}$ /
Rate$_{Low}$), highlighting beginning and end intervals of the 364 ksec
observation. The largest values for the ratio occur at $\sim$8keV -
$\sim$10keV, indicating an overall slight softening trend. Observe that
these spectra are generally considerably steeper than those for the
intense flare activity reported for 2006 \textit{Swift} observations of
Mrk 421 in Tramacere et al. (2009).

In addition to analyzing spectra at the onset and end of the observation, 
we construct 12 smaller time bins describing the entire observation and fit 
the corresponding XIS and PIN spectra with a broken power-law  model. 
The chance probablility associated with the reduced $\chi^{2}$-values of 
the fits have values between 0.01 and 0.96, indicating satisfactory fits. 
The best fit parameters are given as a function of time in Figure \ref{fitbknlot}. 
It is not possible to fit the time averaged spectrum satisfactorily with a power-law, 
power-law with high energy cutoff, or broken power-law models. This is not 
unreasonable due to the wide range of flux and spectral variability observed 
in the shorter time intervals.

\figureoutpdf{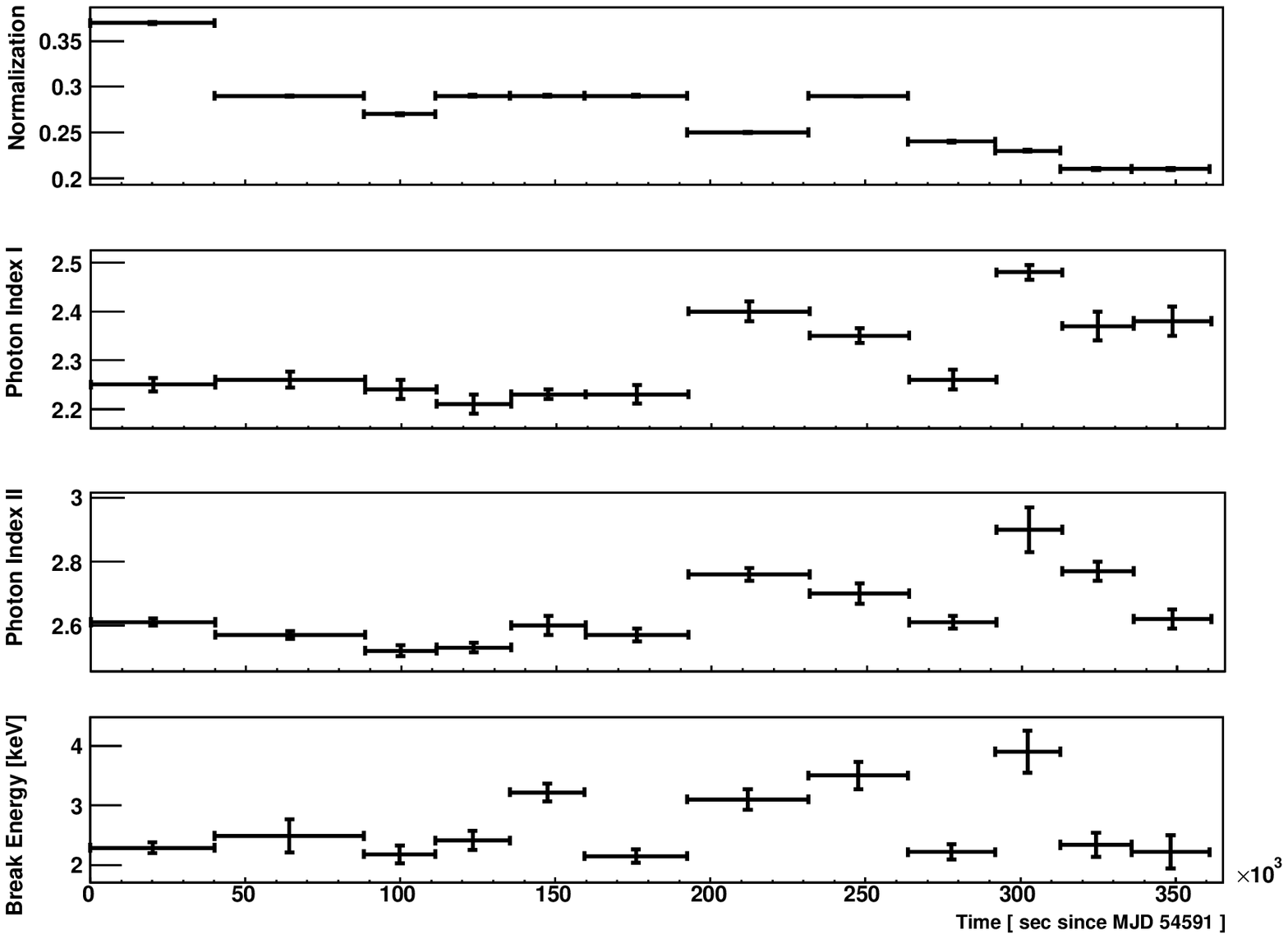}{3.9}{0.0}{0.0}{
%\begin{figure}
%\includegraphics[scale=.8]{figs/fig4.eps}
%\caption{
This figure plots the best fit parameters and 1$\sigma$ confidence 
ranges for different times during the observation using a broken power-law 
model. The  panels (\textit{top-to-bottom}) give the evolution of the flux 
normalization [photons sec$^{-1}$ keV$^{-1}$],   $\Gamma_{1}$,  
$\Gamma_{2}$, $E_{B}$ [keV]. 
  \label{fitbknlot}}
%\end{figure}

Figure \ref{F_PI} investigates the correlations between flux
normalization, $\Gamma_{1}$,  $\Gamma_{2}$, and \teq{\Ebreak} for 
the 12 time intervals in Figure \ref{fitbknlot}. The upper left panel shows
that  $\Gamma_{1}$ increases somewhat for larger  \teq{\Ebreak} values,
which is also the case for  $\Gamma_{2}$ and \teq{\Ebreak}
(\textit{lower left panel}); the scatter in these trends is large. The
upper right panel shows no correlation between \teq{\Ebreak} and the
normalization. Finally, the lower right panel demonstrates that the two
photon indices increase and decrease together, as would be expected with
slight variations in the maximum energy of the radiating particles. 
Insights gleaned from these correlation plots are discussed in
Section~\ref{sec:discuss}.

%\newpage

\figureoutpdf{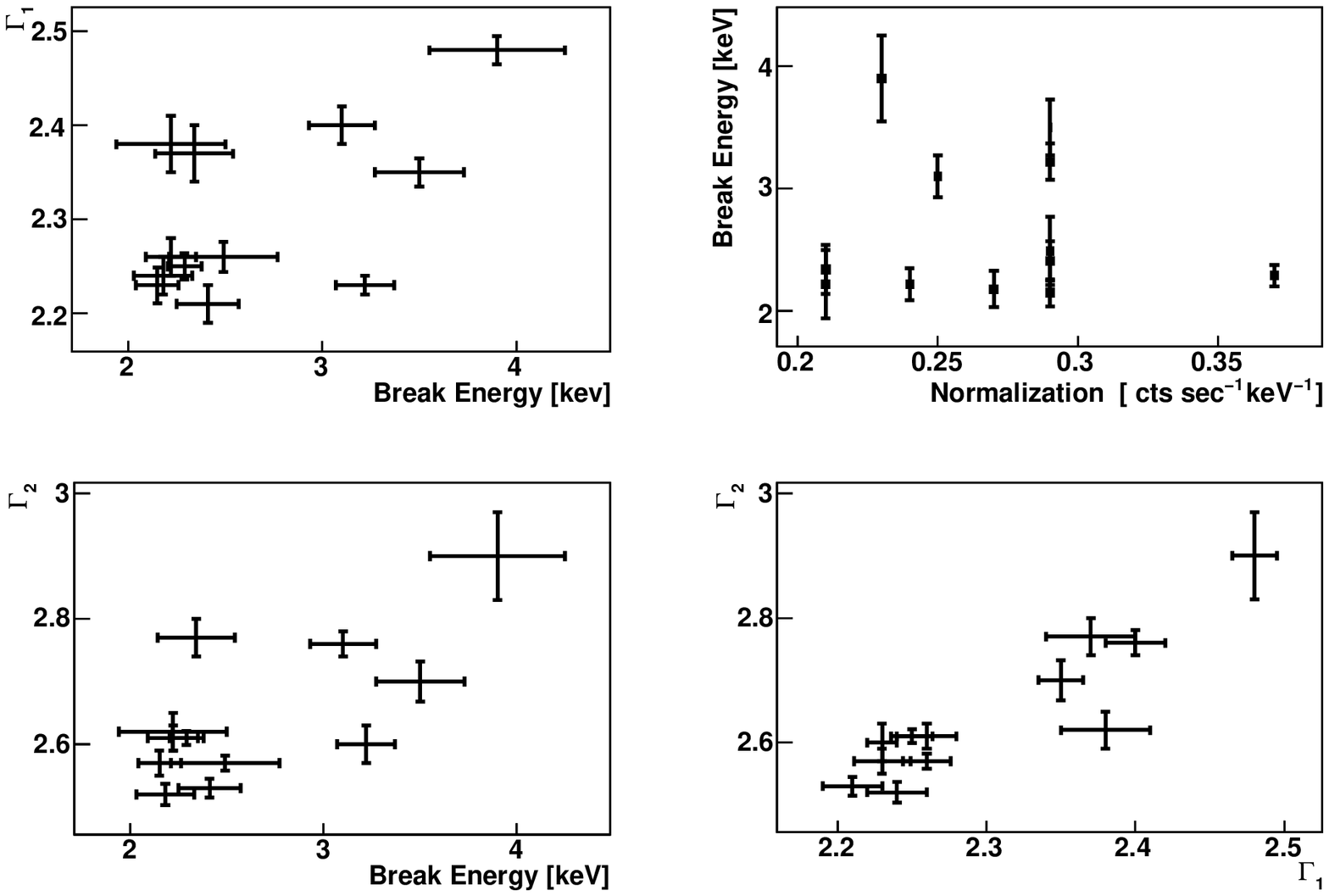}{4.0}{0.0}{0.0}{
%\begin{figure}
%\includegraphics[scale=.8]{figs/fig5.eps}
%\caption{
Best fit parameters from the four panels of Figure \ref{fitbknlot} are 
plotted against each other. The upper left figure plots the correlation of 
\teq{\Ebreak} and  $\Gamma_{1}$. The lower left panel shows the relationship 
between \teq{\Ebreak} and  $\Gamma_{2}$. The upper right panel gives  
\teq{\Ebreak} vs. flux normalization which shows no correlation. The lower right 
panel shows the relationship between  $\Gamma_{1}$ and  $\Gamma_{2}$. 
The two photon indices increase and decrease together.
 \label{F_PI}}
%\end{figure}

Further understanding can be provided by exploring hardness ratio (HR)
variations during flares. In Figure \ref{FVHR}, HR-flux diagrams are
given for the two most prominant flares (1.5$\times$10$^{5}$ sec and
2.5$\times$10$^{5}$ sec Fig \ref{LC}). The left panels show the light
curves in two energy bands from XIS observations. The higher energy
rates have been multiplied by a factor of 3 for clarity. We see that for
the \textit{Flare 1} (\textit{top panels}), there is initial clockwise
motion which quickly changes to a larger counter-clockwise arc through
the HR-flux plane.  \textit{Flare 2} also shows both clockwise and
counter-clockwise motion but in a smaller figure-8 pattern. These
characteristics of spectral hysteresis are similar to a subset of the
\textit{Swift} data reported in Tramacere et al. (2009) for 2006
observations of Mrk 421 flares. The HR-flux trend of an observation can
be an indicator of relative time scales for processes involved with
acceleration and emission (see Kirk, Rieger \& Mastichiadis 1996;
Tramacere et al. 2009 and references therein for a discussion). For
clockwise motion, the cooling time will be longer than the acceleration
time. The two timescales are comparable for counter-clockwise trends.

\section{Spectral Interpretation and Discussion}
 \label{sec:discuss}
One of the key results from this observation campaign is that we observe
different spectral evolution for similar flares. The different spectral
evolutions during different flares exclude models in which flux and
spectral variations are caused exclusively by variations of the Doppler
beaming factor. Furthermore, they do not concur with simple models where
particles are always accelerated with the same spectral index and cool
radiatively. Therefore, we conclude there have to be intrinsic
variations of the spectral index and density of the radiating (electron)
population. To be more precise, such particles are injected with a
power-law $dN_e/d\gamma_e \propto n_e\gamma_e^{-\sigma}$, and both 
$n_e$ and $\sigma$ vary from flare to flare and also during a flare. This
conclusion is underpinned by our reported \textit{Suzaku} detection of
rapid flux and index variations in the somewhat steep X-ray spectra.

\twofigureoutoverpdf{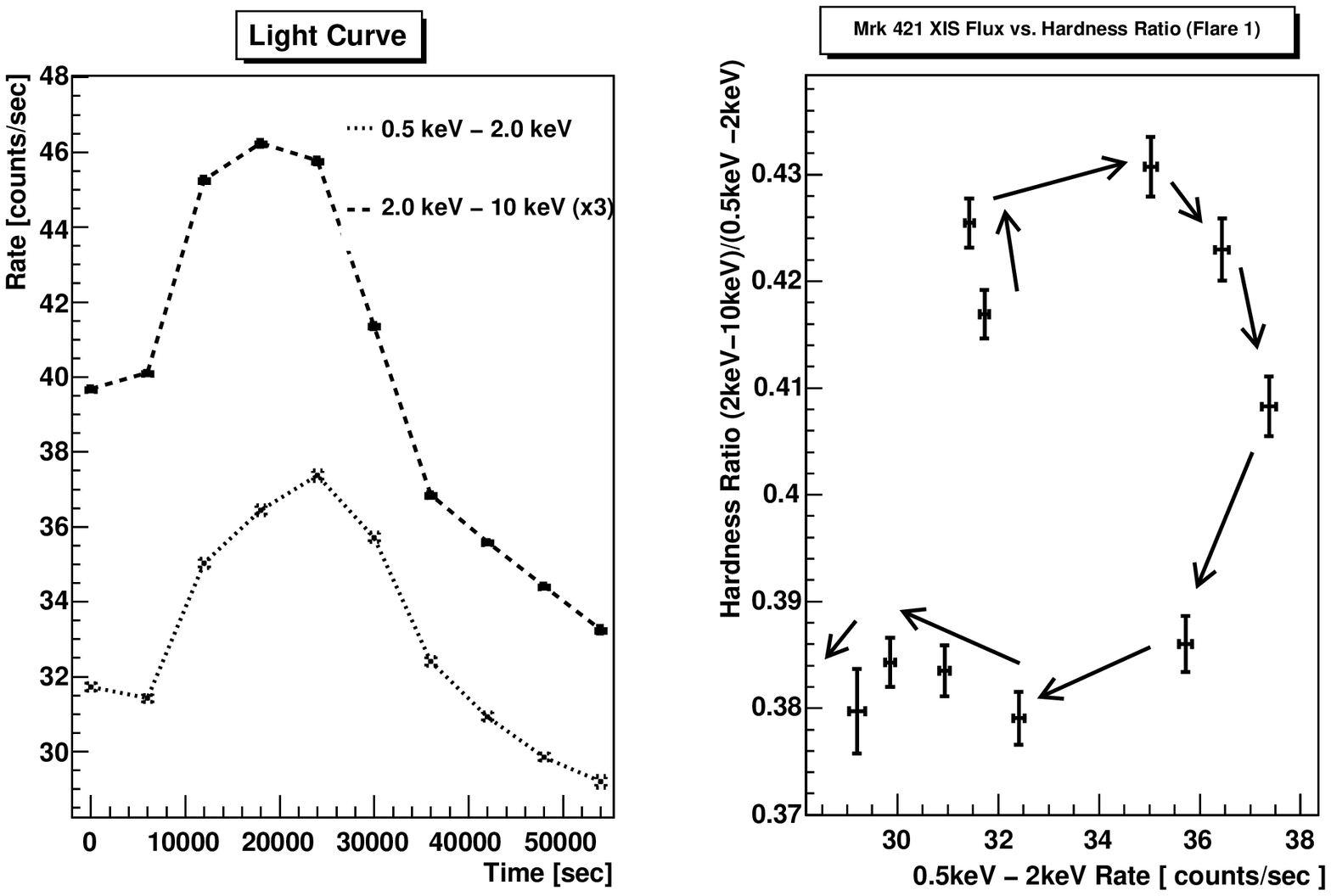}{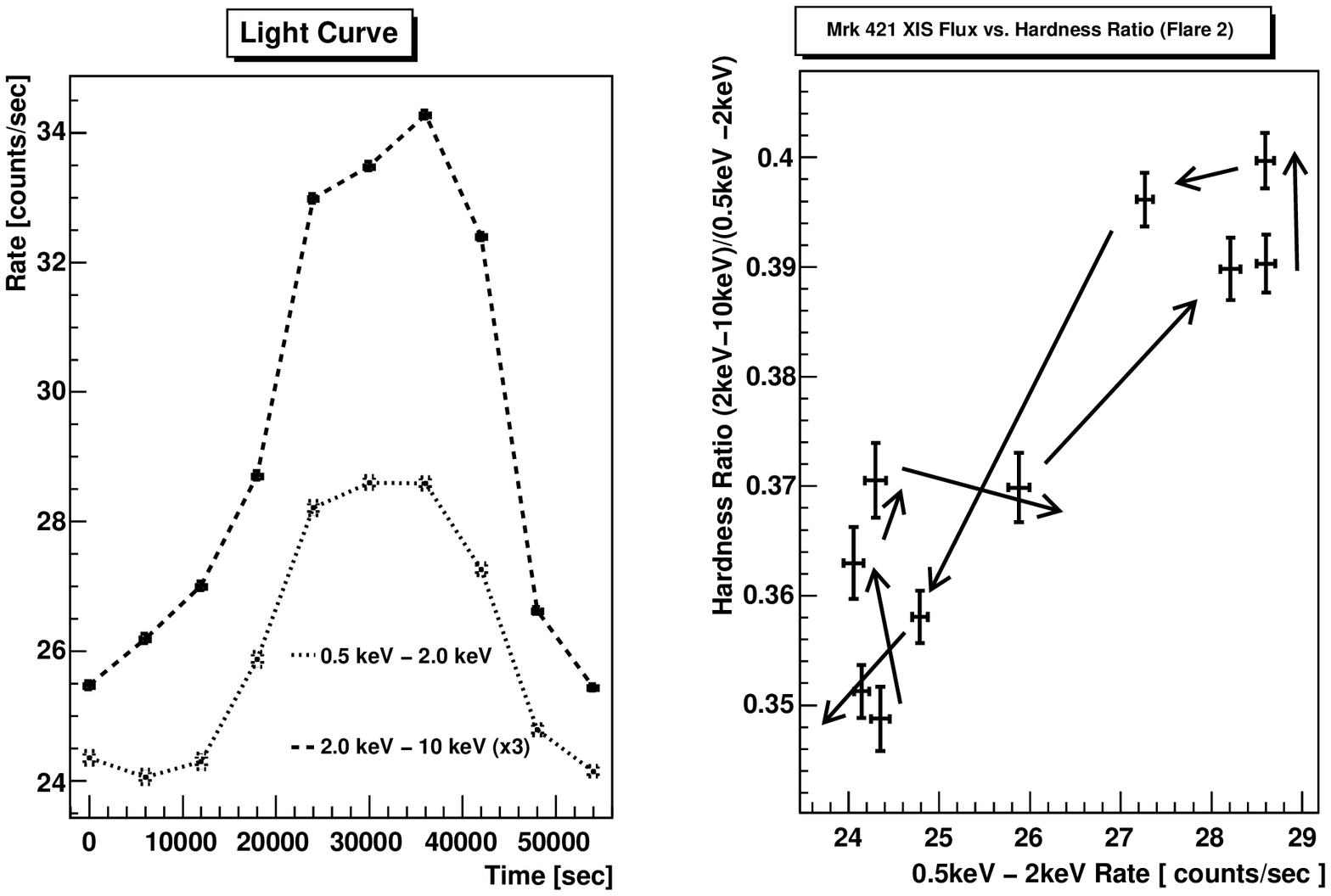}{
%\begin{figure}
%\includegraphics[scale=.55]{figs/fig6a.eps}\\
%\includegraphics[scale=.55]{figs/fig6b.eps}
%\caption{
The flux vs. hardness ratio plots of two well-defined flares 
show different motions in the HR-flux plane. The left panels show the 
XIS light curves for the flares in two energy bands: 0.5 keV - 2 keV 
(\textit{$Rate_{high}$: dotted line}) and 2 keV - 10 keV 
(\textit{$Rate_{low}$: dashed line}). The rates in the higher energy 
band have been multilpied by 3 for clarity. The right panels plot the 
hardness ratio ($Rate_{high}$/$Rate_{low}$) vs $Rate_{low}$ correlation 
for these flares.  For \textit{Flare 1} (\textit{upper right}), there is initially 
a clockwise trend followed by larger counter-clockwise movement. The 
lower right panel also shows a shift from clockwise to counter-clockwise 
but in a figure 8 pattern. Here, \textit{Flare 1} (\textit{top panels}) occurs at 
$\sim$1.5$\times$10$^{5}$ sec into the observation and \textit{Flare 2} 
(\textit{bottom panels}) occurs at  $\sim$2.5$\times$10$^{5}$ sec.
 \label{FVHR}}
%\end{figure}

\vspace{0.05truein}

In the following interpretative discussion, two scenarios will be
addressed in turn, after the issue of Doppler boosting variations is
first touched upon. First, the case that radiative cooling does not lead
to a steepening of the observed X-ray energy spectra is considered.
Subsequently, we turn to the scenario of efficient radiative cooling
that spawns a steepening of the observed spectral indices by $\delta
\Gamma=1/2$.  We remark that several VERITAS TeV gamma-ray 
observations were taken during our \textit{Suzaku} observations 
(see Acciari et al. 2009; Acciari et al. 2010).  These
observations revealed modest TeV gamma-ray flares, demonstrating that
the synchrotron spectral power dominated the inverse Compton emissive
power, which is also the situation for much earlier Whipple-era
observations (e.g. see the broadband data depiction in Inoue \& Takahara
1996). The following discussion can thus safely neglect complications
arising from inverse Compton cooling in the Klein-Nishina regime, which
in other circumstances can modify the distribution of the highest energy
particles, and therefore also the shape of the X-ray synchrotron
continuum.

To provide context for the results of this paper in the light of other
multifrequency observations, we note that it is likely that the radio
emission observed from Mrk421 does not come from precisely the same
spatial region as the X-ray emission. For BL Lac type objects, no
convincing radio/x-ray correlation has ever been established. For
leptonic models that explain the observed X-ray emission without giving
a measurable radio flux, see Rebillot et al. (2003) and Krawczynski et
al. (2001). The discussion here therefore focuses on explaining the high
energy emission from electrons close to the high energy cutoff of the
electron energy spectrum. While multiwavelength SED modeling of blazars
with radio-to-X-ray synchrotron and gamma-ray inverse Compton signals
typically constrains the approximate maximum Lorentz factor of the
electrons, the mean magnetic field strength and the bulk Doppler factor
\teq{\delta} of the jet (e.g. see Bednarek \& Protheroe 1997;
Mastichiadis \& Kirk 1997), SED fluctuations augment such by probing
different jet environmental quantities.  This is the focus here, and
X-ray observations afford stronger diagnostics than do the GeV and TeV
bands due to their high count rates. For example, typical variability in
{\it Fermi}-LAT data on blazars samples timescales of a few days to a
week at best (Abdo et al. 2010b) when acquiring sufficient count
statistics. TeV gamma-ray energy spectra with spectral index errors
$<0.1$ can be acquired on short ($\sim$10 min) time scales - but require
extremely strong flares.

To begin the interpretive focus on the X-ray variability, in principal,
adjustments of the spectral indices and hardness ratios can be generated
by fluctuations in the Doppler beaming factor $\delta$ during flares. 
Such \teq{\delta} variations can arise in bent jets, such as the
scenario envisaged for {\it Fermi}-LAT and other wavelength observations
of the quasar 3C 279 (Abdo et al. 2010a).  Consider the correlation
plots in Figure~\ref{F_PI}. In the absence of other influences, changes
in $\delta$ should manifest themselves as a scaling of the break energy
\teq{\Ebreak\propto \delta} (blueshifting) {\it and} an associated
scaling of the flux at $E_{\rm br}$ as $\delta^4$.  This correlation
should hold approximately even if the 0.2-10 keV window samples a
portion of a larger SED curvature.   It is clearly not seen in the upper
right panel of Figure~\ref{F_PI}, indicating that some other
environmental fluctuation is operative.  If the SED curvature is broad,
one might expect that higher break energies in the fits will correlate
with lower $\Gamma_1$.  The opposite is suggested in the upper left
panel of  Figure~\ref{F_PI}, but the scatter is large, and the
\teq{\Ebreak} range is small. In terms of the HR-flux diagrams in
Figure~\ref{FVHR}, pure Doppler factor $\delta$ fluctuations should
yield a strong correlation between the hardness ratio and the count rate
below 2 keV: essentially a diagonal trace from lower left to upper
right.  This is clearly not seen for \textit{Flare 1}.  There is more of
an indication for \textit{Flare 2}, but significant deviations from a
clean correlation occur when taking into account the evolutionary track.
Accordingly, something deeper than just simple Doppler boosting
variations must be active in the jet environment, and our attention
turns to the particle acceleration properties in shocks within the Mrk
421 jet.

The observed spectral fluctuation signatures are therefore interpreted
now in the light of expectations from diffusive shock acceleration
theory. This logical step can be taken because the power-law is both
well-established below \teq{\Ebreak} and is considerably flatter than
the steeper spectra seen in the TeV band that might typify the onset of
a cutoff. If the spectral indices $2.2 \lesssim \Gamma_1 \lesssim 2.5$
identified in Figures~\ref{F_PI} and~\ref{FVHR} are attributed to
synchrotron emission from a particle distribution $dN_e/d\gamma_e
\propto \gamma_e^{-\sigma}$ (below a maximum Lorentz factor cutoff
$\gammax$), then for uncooled synchrotron scenarios, $\Gamma_1 = (\sigma
+1)/2$. These cases are where the accelerated population is continually
replenished in the emission region on timescales inferior to the
synchrotron cooling timescale $\tsyn = 4\pi m_ec/(\sigma_T\, B^2)\,
\gamma_e^{-1}$. Observe that \teq{\tsyn\sim 1}hour for \teq{B=0.1}G and
\teq{\gamma_e\sim 10^7}, parameters that would place the synchrotron
turnover at \teq{\sim 150 \delta}keV.  In this scenario, the
acceleration timescale needs to be comparable to the flare duration or
shorter, while \teq{\tsyn} needs to exceed the flare timescale, a
situation that occurs for lower \teq{\gamma_e} that can move the
synchrotron turnover down to the \textit{Suzaku} window.  For such an
uncooled acceleration picture, the X-ray index in any time interval
leads to a constraint on the electron index $\sigma$, and we find
$3.4\lesssim\sigma \lesssim 4$.  The physical conditions in the Mrk 421
jet environment that can generate $\sigma$ in this range can be assessed
using the Monte Carlo simulational modeling of Baring \& Summerlin
(2009), Baring (2010), and Ellison \& Double (2004), of
particle acceleration at relativistic shocks. These works provided a
useful and expansive complement to earlier semi-analytic investigations
of Kirk \& Schneider (1987) and Kirk \& Heavens (1991) that employed
eigenfunction techniques to solve the diffusion-convection equation at
mildly-relativistic shocks.

The spectral index parameter space explored in these simulational
studies clearly indicated that values of $\sigma > 3$ are appropriate
only for so-called {\it superluminal shocks}, i.e. those where
$u_1/\cos\Theta_{Bn1} > c$.  Here $u_1$ is the component of the upstream
flow speed normal to the shock in its rest frame.  For the relativistic
outflows commonly invoked in blazar jets, one naturally expects
$u_1\sim$c.  Also, $\Theta_{Bn1}$ is the angle the magnetic field vector
makes to the shock normal in the upstream fluid rest frame.  Therefore,
superluminal (and oblique, $\Theta_{Bn1} > 0^{\circ}$) conditions in Mrk
421's jet would naturally be expected.  However, Baring \& Summerlin
(2009) also observed that to generate $\sigma > 3$, it would be
necessary for the field turbulence in the shock neighborhood to not be
too strong, perhaps limiting field fluctuations to $\delta
B/B\lesssim0.1$, so that particle diffusion is not too near the
isotropic Bohm limit (essentially occurring for \teq{\delta B/B\sim 1}).
This is an interesting environmental constraint that lowers the expected
acceleration time \teq{\tacc} (e.g. Jokipii 1987) due to the inefficient
trapping of charges in oblique shocks, a property that is directly
responsible for steeper power-laws with $\sigma \gtrsim 3$.

Consider instead a strongly-cooled synchrotron emission picture, where
the shock injects relativistic particles into a larger region where the
synchrotron cooling timescale exceeds the injection timescale.  Invoking
such to explain the \textit{Suzaku} power-law indices, one infers
$\sigma = 2(\Gamma_1 -1)$ for the shock acceleration spectral index,
since synchrotron cooling steepens the electron power-law by an index of
unity; the index then falls in the range $2.4\lesssim\sigma\lesssim 3$
for the data reported in Figures~\ref{F_PI} and~\ref{FVHR}. This still
lies in the parameter space for superluminal shocks (Baring \& Summerlin
2009), but requires somewhat stronger field turbulence than for the
uncooled case, perhaps in the range $\delta B/B\lesssim 0.3$.  Again
Bohm-limited diffusion is not indicated.  It is noted in passing that
these claims are predicated on acceleration theory results generated for
small angle scattering (i.e. pitch angle diffusion); if $\delta
B/B\sim1$ is considered, then one anticipates that larger angle
deflections of charges will be active, resulting in much flatter spectra
(e.g. Ellison, Jones \& Reynolds 1990; Stecker, Baring \& Summerlin
2007) that are incongruent with the Mrk 421 \textit{Suzaku} data
presented here.  Such a large angle scattering regime may be more
appropriate for the 2006 \textit{Swift} observations of the intense
flare in Mrk 421 (Tramacere et al. 2009), and for some flat spectrum
gamma-ray sources in the {\it Fermi}-LAT database (Abdo et al. 2009).

For Mrk 421, 2009 observations by {\it Fermi}-LAT that are not
contemporaneous with the \textit{Suzaku} data presented here yield
$\Gamma_{\gamma}\sim 1.78$ (Abdo et al. 2009).    If this signal
constitutes inverse Compton emission by uncooled electrons at Lorentz
factors below \teq{\gamma_c} (\teq{<\gammax}), then one infers
$\sigma\sim 2.56$, not dissimilar from the \textit{Suzaku} inference for
strong cooling just above.  Given that this GeV-band spectrum probably
originates from electrons that emit synchrotron photons below the X-ray
window, and that the steeper TeV spectrum (\teq{\Gamma_{\gamma}\sim
2.91} in the contemporaneous VERITAS data presented in Acciari et al.
2009) provides an approximate inverse Compton image of the X-ray
synchrotron signal (with\teq{\Gamma_2\sim 2.5 - 2.9} here), one expects
the inferred \teq{\sigma} for the {\it Fermi} data should be slightly
lower than that for the \textit{Suzaku} observations.  Note also that
historically, the radio spectrum for Mrk 421 is flatter still, at
\teq{\Gamma_{\hbox{\sevenrm rad}}\sim 1.1 - 1.3} (e.g. see Makino, et
al. 1987), suggesting \teq{\sigma\sim 1.2 - 1.6} for the electrons
radiating at these frequencies.  Taken together with the gamma-ray data,
a picture emerges that the radiating lepton distribution might be
injected with a ``convex'' distribution, i.e. with \teq{\sigma} an
increasing function of energy.  Yet, care must be taken to explore the
influence of non-cospatiality for the origin of the various emission
components, and the role of synchrotron self-absorption, before
diagnosing such a curvature in the injection distribution.

Let us delve deeper into a comparison between the cooled and uncooled
emission scenarios. It is possible to envisage a cospatial competition
between acceleration and synchrotron cooling, a paradigm that is
commonly accepted in models of X-ray emission in Galactic supernova
remnants (SNRs).  While this can generate the observed variability in
both flux and spectral index, unless diffusion in shock-layer turbulence
is incredibly inefficient, the requirement that a cooling-limited
synchrotron turnover fall in the \textit{Suzaku} X-ray window constrains
the shock speed $u_1$ to values around $0.01c$, independent of the
strength of $B$, provided that the acceleration process is gyroresonant,
which is the prevailing paradigm. This assertion can be justified using
results from the discussion of cooling-limited SNR shock acceleration in
Baring et al. (1999). Eq.~(12) therein indicates that the acceleration
rate gives $d\gamma_e/dt\propto (u_1/c)^2\, eB/(\eta\, mc) $, where the
ratio $\eta =\lambda/r_g\geq 1$ of the particle's mean free path
$\lambda$ to its gyroradius $r_g$ measures the departure from isotropic
Bohm diffusion ($\eta=1$, i.e. $\delta B/B\sim 1$). This can be equated
to the synchrotron loss rate $\vert d\gamma_e/dt \vert \propto
\gamma_e^2 B^2$ in the comoving frame of the jet. The resulting electron
Lorentz factor $\gamma_e\equiv\gamma_c\propto u_1 (\eta B)^{-1/2}$ for
cooling-limited acceleration can be inserted into the textbook formula
for the characteristic synchrotron energy to yield a synchrotron
peak/cutoff energy, that is independent of the field strength:
\begin{equation}
  E_{\rm syn} \;\sim\; \frac{\delta}{\eta}
     \left( \frac{u_1}{c} \right)^2 \, \frac{m_ec^2}{\alpha}\quad 
\end{equation}
Here $\alpha=e^2/(\hbar c)$, and the blueshift due to Doppler beaming
has been included. For $\delta =1$, the $u_1=c$, $\eta=1$ limit of this
is around 50 MeV, as was highlighted in De Jager et al. (1996) for
considerations of gamma-ray emission at relativistic pulsar wind nebular
shocks; see de Jager \& Baring (1997) for a compact presentation of this
critical energy.

To move $E_{\rm syn}$ into the classic X-ray band one has to set
$u_1\sim 0.01c$ if  $\eta\gtrsim 1$ and lower still if $\delta > 1$.
This lower shock speed is an attractive value for SNRs, but is clearly
too small for blazar jet contexts. It is possible to adjust $\eta$ to
fix $u_1\sim c$, which quickly leads to fitting values $\eta\sim 10^5$,
thereby dramatically reducing the rapidity of the acceleration process. 
This was the approach of Inoue \& Takahara (1996) when exploring
multiwavelength modeling of Mrk 421 spectra (they required even higher
values $\eta\sim 10^7$ for their 3C 279 case study), who assumed
$\delta\sim 10$. In the light of refined studies of acceleration at
relativistic shocks, this is unsatisfactory on three counts.  First, the
parameter space of shocks that would generate indices $\sigma$ that
would accommodate the \textit{Suzaku} indices is extremely constrained
to the subluminal/superluminal boundary (e.g. see Baring 2010).  Next,
requiring $\eta > 10^4$ leads to extraordinarily inefficient injection
of particles into the acceleration process (e.g. Baring \& Summerlin
2009), imposing uncomfortable constraints on blazar energetics. 
Finally, such large values of $\eta$ define essentially laminar fields
that are not expected in shocks, which are inescapably turbulent. 
Hence, it is difficult to fine-tune a synchrotron-cooling
limited-turnover in the X-ray band in the blazar model context.

In contrast, it is quite possible that a cooling break can be situated
below the X-ray band, provided that the acceleration and cooling regions
are spatially distinct. This is a preferred paradigm in many blazar
models. Such a strongly-cooled case corresponds to static or impulsive
acceleration at a shock, generating non-thermal electrons up to the
maximum Lorentz factor (which can be \teq{\gamma_e\gtrsim 10^6} on
timescales of a few seconds for \teq{\eta\sim 1} and \teq{B\sim 0.1}G),
followed by escape from the shock environs and subsequent gradual
cooling in a remote and more extended region that is defined by the
competition between spatial diffusion/convection and radiative cooling.
Flux and index variability driven by cooling effects would then tend to
be muted by spatial and temporal convolutions.  Moreover, spectral
cooling breaks, if situated in the optical/UV band, would correspond to
Lorentz factors \teq{\gamma_e \sim 10^4 - 10^5} and therefore yield
cooling times of the order of days or longer.  Hence the spectral
variations on timescales of a few hours reported here very probably
reflect intrinsic fluctuations in the acceleration/injection process, as
opposed to spatial inhomogeneities such as magnetic field clumping in
the cooling region.  The spectral hysteresis evinced in
Figures~\ref{F_PI} and~\ref{FVHR} possesses some similarities to, and
significant differences from that envisaged in the competitive
acceleration/cooling model of Kirk et al. (1998).  \textit{Flare 1}
seems to suggest that alterations in shock conditions precipitate an
increased injection rate \teq{{\dot n}_e} (or an increased field) {\it
before} flattening the distribution (lowering index \teq{\sigma}), the
system subsequently relaxing via reducing the injection rate or field
strength and finally displaying signs of an incipient increase in
\teq{\sigma}. Field turbulence variations should drive injection and
\teq{\sigma} changes that contribute to both flux and hardness ratio
alterations.  \textit{Flare 2} encapsulates another level of complexity,
defying simple description.

To summarize, given these cooling/acceleration considerations, it seems
likely that the variations depicted in Figures~\ref{F_PI} and~\ref{FVHR}
signify changes in the lepton acceleration at the relativistic shocks
contained in the Mrk 421 jet, perhaps with a smaller contribution from
Doppler beaming fluctuations. The acceleration fluctuations are easily
produced from a theoretical standpoint by just modest or small changes
to the level of field turbulence, the mean field direction and
amplitude, or the local density encountered in the shock environs as it
traverses jet material.  This claim is underpinned in part by the
broad-ranging spectral index phase space plots presented in Baring \&
Summerlin (2009) and Baring (2010), together with their discussion of
correlated injection efficiencies.

Finally, even though properties of the acceleration process may cause
spectral variations in the X-ray band, we reiterate that most of the
conclusions from earlier multiwavelength leptonic modeling work are
still valid (e.g. Krawczynski et al. 2001). A magnetic field of
$\sim$0.2 G is still needed so that electrons can emit a good fraction
of their energy on $\sim$1 hr time scales.  Also, $\gammax$ and
$\delta$/B are still contrained by the relative peak positions of the
SEDs in the x-ray and gamma-ray regimes. The inference of variations of
the parameters of the acceleration process in jet shocks from observed
X-ray SED fluctuations is a subtlety that does not substantially modify
these more global parameters inferred from the broadband SED modeling,
but does directly impact the relative apportionment of acceleration and
cooling, in part through constraints imposed on \teq{\eta =\lambda/r_g}.

\section{Summary}

We present the results from a 4-day \textit{Suzaku} observation of Mrk
421 while in a flaring state. The 0.5 keV - 30 keV flux decreased by a
factor of 2 during the course of the observation. We find good agreement
when fitting spectra from isolated time intervals with a galactic
absorption + broken power-law model. Trends in the HR-flux plane
indicate there are different timescales for competing processes which
differ from flare to flare and for different flux levels.  The X-ray
spectral index changes by $\sim$0.2. However, the spectral evolution
seems not to be related to the phase of a flare.  The erratic relation
between the light curves and the spectral indices suggests constraints
on the interpretation of the shocked jet environment.

In the literature, the past observations have often been explained by
invoking the competition of the acceleration and cooling time scales.
However, it seems improbable that the timescales of acceleration and
cooling are similar, since this would require jet shock speeds of the
order of $0.01c$, and that the relative importance of shock acceleration
and subsequent synchrotron cooling differs from flare to flare. We
suggest here that it is more likely that the \textit{Suzaku} data
properties reported here are due to intrinsic changes in the
acceleration process at relativistic shocks in the jet, producing
electron distributions with varying spectral indices and changing
maximum energies.

\vspace{0.2truein}

\acknowledgments
This research has made use of data obtained from the \textit{Suzaku}
satellite, a collaborative mission between the space agencies of Japan
(JAXA) and the USA (NASA). AG and HK acknowledge support by the Suzaku
Guest Investigator Programme, NASA grant NNX08AZ76G.  MGB acknowledges
support from National Science Foundation grant PHY07-58158 and NASA
Astrophysics Theory grant NNX10AC79G.  MGB is also grateful to the Kavli
Institute for Theoretical Physics, University of California, Santa
Barbara for hospitality during part of the period when this research was
performed, a visit that was supported in part by the National Science
Foundation under Grant No. PHY05-51164.

\end{document}